\documentclass[preprint]{aastex}
\usepackage{epsfig}

\title{Canonical Particle Acceleration in FRI Radio Galaxies}

\author{Andrew Young\altaffilmark{1,2}, Lawrence Rudnick\altaffilmark{1}, Debora Katz
\altaffilmark{3}, Tracey DeLaney\altaffilmark{1,4}}
\author{  Namir E. Kassim\altaffilmark{5} and 
Kazuo Makishima\altaffilmark{7} }

\altaffiltext{1}{Department of Astronomy, University of Minnesota, 116 Church 
Street SE, Minneapolis, MN  55455}
\altaffiltext{2}{Department of Physical Science, Kutztown University, Kutztown, PA 19530  }
\altaffiltext{3}{United States Naval Academy, 121 Blake Rd., Annapolis, MD 21402}
\altaffiltext{4}{Harvard-Smithsonian Center for Astrophysics, 60 Garden Street MS-67, Cambridge, MA 02138}
\altaffiltext{5}{ Remote Sensing Division, Naval Research Laboratory, 
Washington, DC 20375-5320}
\altaffiltext{6}{Department of Physics, School of Science, The University of Tokyo, 7-3-1 Hongo, Bunkyo-ku, Tokyo 113-0033, Japan}

\date{\today}

\begin{abstract}
Matched resolution multi-frequency VLA observations of four radio
galaxies are used to derive the asymptotic low energy slope of the
relativistic electron distribution. Where available,  low energy
slopes are also determined for other sources in the literature.  They
provide information on the acceleration physics independent of
radiative and other losses, which confuse measurements of the
synchrotron spectra in most radio, optical and X-ray studies. We find
a narrow range of inferred {\it low energy} electron energy slopes
n(E) $\propto$ E$^{-2.1}$ for the currently small sample of 
 lower luminosity sources classified
as FRI ({\it not classical doubles}).  This distribution is close to, but
apparently inconsistent with, the test particle limit of n(E) $\propto$ E$^{-2.0}$
expected from strong diffusive shock acceleration in the
non-relativistic limit. Relativistic shocks or those modified by the
back-pressure of efficiently accelerated cosmic rays are two
alternatives to produce somewhat steeper spectra. We note for further
study the possiblity of acceleration through shocks, turbulence or
shear in the flaring/brightening regions in FRI jets as they move away
from the nucleus. Jets on pc scales and the collimated jets and
hot spots of FRII (classical double) sources would be governed by
different acceleration sites and mechanisms; they appear to show a
much wider range of spectra than for FRI sources.

\end{abstract}


\begin{document}

\section{Introduction}
 \label{intro}
The acceleration of relativistic particles in radio galaxies occurs at
multiple sites, and probably through multiple mechanisms.  As we have
long known, particles are first accelerated to relativistic energies
near the nucleus, where they illuminate jets and other structures on
pc scales \citep{coh77,jon74}.  At scales of $10^2$ - $10^3$ pc, {\it
in situ} acceleration must also be present in some sources, in order
to produce the short radiative lifetime relativistic particles seen in
optical and X-ray synchrotron radiation (see review by
\citet{wor04}). Finally, $10^4$ to $10^5$ pc away from the nucleus,
powerful radio galaxies often contain compact radio and even optical
``hot spots'', yet another site of relativistic particle acceleration
\citep{mei+89} powered by the flow from the AGN (see
\citet{gop01,aha02} for alternative views).

In this paper, we investigate one key aspect of the acceleration
processes, {\it viz.}, what is the limiting logarithmic 
slope of the relativistic
electron energy distribution at low energies.  This slope is an
indicator of the basic acceleration processes, although it represents
the full acceleration history of the electrons through a variety of
sites.  At higher energies (higher observed synchrotron
frequencies), the confusing effects of radiative and other
losses set in and the original signatures of the acceleration process
are usually obscured. At sufficiently low energies, the power law
may be affected, e.g., by ionization losses, but no such effects
are visible in the data presented here.

For practical reasons, most of the information available on low
frequency energy slopes is for Fanaroff-Riley I (FRI) sources, those
without bright peaks of emission at their ends \citep{fan74}.  This
class includes the tailed radio galaxies as well as more diffuse
sources, and they typically have low radio luminosities
( $\lessapprox 10^{24.5}$ Watts/Hz at 1.4 GHz, dependent on the host
optical luminosity \citep{owe94}).  We also include some brief
references to the information available on FRII sources, which
include the prototypical ``classical doubles''.

\subsection{Low energy/frequency slope measurement challenges}

Although the importance of the low energy slope for understanding
particle acceleration has been recognized for decades, these
measurements are not easily done.  Historically, integrated spectra
have been assumed to reflect the low frequency power law
(e.g. \citet{alex87}), but the two values are virtually unrelated as
discussed further below.  A second method has been to fit
multi-frequency spectra with standard spectral models \citep{car91}.
However, the models often provide poor fits to the data, and the
low-frequency index is either allowed to float as a function of
position (which would indicate fresh relativistic acceleration at each
location) or is assumed to be some fixed value
\citep{alex84}.

In order to perform spectral measurements, maps at the different
frequencies must be made at the same sufficiently high resolution.  This may result in
beam sizes that are larger than spectral variations in the source, as
well as the more pernicious effects of variations along the line of
sight.  Such variations arise from a number of structures and
processes in radio galaxies.  FRI jet spectra are often contaminated
by the presence of surrounding sheaths that have locally steeper
spectra. These can be detected through tomographic analysis
\citep{kat97,kat+99, hard99} or through polarization structure
\citep{swa98}.  FRII jets should sample the material emerging from the
nucleus, but their spectra are often difficult to determine because
they are embedded in locally steeper spectrum lobes \citep{tre+01}.
The lobes themselves are difficult to use for determining the original
low frequency index.  As seen in the simulations of FRII sources by
\cite{tre01,tre04}, populations of lobe  electrons with very different histories
may be found along any given line of sight. A few will have passed
through a strong terminal shock; many will have experienced
acceleration in the network of weaker shocks in the hot spot region.

The general effect of these non-uniformities on the observed spectrum
 is to broaden it toward a power law, even if every volume element
 contained an exponentially cutoff electron distribution. The effect
 is worst for integrated spectra, which sum over all the emission in
 the source. As shown by \cite{you+02}, the resulting approximate
 power law is not simply related to the low frequency spectral index,
 and so integrated spectra are not reliable indicators of the original
 relativistic particle acceleration as mentioned above.  Lower levels
 of contamination tend to preferentially {\it flatten} the steepest
 spectral indices for simple arithmetic reasons.  Curiously, this
 can have the effect of {\it steepening} the derived low frequency index;
 this occurs because the flattening is more prominent at higher
 frequencies, and the reduction in spectral curvature forces the fits
 to assume that the observations are thus closer to the low frequency
 limit.

\subsection{Methods for measuring low frequency slope}

 In this paper, we invoke three methods to isolate the low frequency
 spectral slope. In each case, we either avoid or take care to
 minimize problems with integration along the line of sight.  The
 first method is the most direct, viz. that spatially resolved spectra
 are available at no less than three frequencies in the low-frequency
 asymptotic power law portion of the spectrum.  This is rarely
 available, and only one example from the literature, M87, is used
 here.

  The second method  is the use of color-color (CC) diagrams.  These are based on
the {\it ansatz} that there is a single shape to the
relativistic electron population throughout the source.  The spectral
{\it shape} is not affected by adiabatic or radiative losses or
magnetic field strength variations; the observed two-point spectral
indices are, of course, affected by these changes. Color-color diagrams provide an equivalent mapping
of the $(log I, log \nu)$ spectrum, but allow one to put all the
data from different positions in the source on one plot.  Color-
color diagrams effectively measure the local curvature of the
spectrum ($\alpha_{\lambda_1}^{\lambda_2} - \alpha_{\lambda_2}^{\lambda_3}$)
at different places along the spectrum (as characterized by 
$\alpha_{\lambda_1}^{\lambda_2}$).  This results from the shift of the
spectrum from one location to another due to changes in local magnetic
fields or radiative or adiabatic losses.  The assumption of
a single electron population is tested by looking for more than a
single locus of points in the CC diagram, with a shape given by one of
the standard models \citep{kat93}.  If an appropriate single locus is
seen, then we perform a one parameter fit for the low frequency
index using a standard model.  This procedure works even if the
low frequency index is not directly observed, although the
errors will, of course, increase along with the required extrapolation.  
In the complicating case where multiple
particle populations are present within each beam, the procedure will visibly fail,
as noted further below.

 In this paper, we perform least-square fits to the CC data by
using the Jaffe-Perola (JP, \cite{jp} model
which describes synchrotron loss spectra when the pitch angles of the
radiating electrons are continuously re-isotropized. These give very
similar low frequency slopes to fits using the alternative
Kardashev-Pacholcyzk \citep{kard,pach} spectral shape, which falls off more
slowly, as long as the observed range in two-point spectral indices is
$\lessapprox1$.  If the two-point spectra change by more than that
across the source, then only the JP models will fit.

The third method for determining  the low-frequency spectral slope is
to use spectral tomography to isolate the spatial trends in jet
spectral index (3C449: \citet{kat97};  1231+674 and
1433+553: \citet{kat+99}). Tomography simply involves the subtraction
of images at two frequencies with a variety of scaling factors. One
often observes the disappearance of a distinct structure (e.g., a jet) in
one of these residual maps, with flux from a second component (e.g., a
diffuse sheath) still present at the same location. In such cases, the
spectral index derived from the two original maps is some mixture from
the two overlapping components; the scaling factor responsible for a
structure's disappearance, however, directly gives the structure's
spectral index.    The
asymptotic spectral index observed as the jets approach the
nucleus  is most likely the low frequency power law.
If instead, the underlying electron spectrum were actually curved in
this region, then the observed local changes in jet brightness (including
magnetic field) would have led to changes in observed spectral index between
two fixed frequencies. 
 Without tomography,  regions of constant spectral
index are very difficult to detect because of confusion from the
steeper spectrum sheaths.

\subsection{Acceleration mechanisms}

A good review of possible particle acceleration mechanisms can be
found in \citet{eil91}.  The most popular scheme is Fermi
acceleration \citep{fer49}; in the case of first-order Fermi
acceleration, (diffusive shock acceleration) particles cross a shock
boundary where they can gain energy through head-on collisions
\citep[e.g.,][]{eil91, dru89, kir89}. Relativistic shocks
\citep{ach01} and shocks that are modified by the back-pressure of the
accelerated cosmic rays \citep{bar91} change the shape of the
synchrotron spectrum and its low frequency slope.  Other acceleration
mechanisms rely on a direct current electric field. In these
cases, the energy required may come from magnetic field line
reconnection \citep[e.g.,][]{eil91, lit99,les98}. Other mechanisms
involve the gross kinematics of the plasma; for example, \citet{rie02}
(and references therein) describe shear and centrifugal acceleration
of particles.  In general, these ``second order" processes are less
efficient than diffusive shock acceleration.  In all cases, the low
frequency slope of the relativistic particles approaches a power law
that reflects the acceleration physics \citep{bla87,ost93}. In the case of
test particle diffusive shock acceleration, the slope is simply
related to the Mach number of the accelerating shock.  For a power law
distribution of electrons, N(E)~$\propto~$E$^{s}$ there is a power law
synchrotron spectrum, I($\nu$)$\propto \nu^{\alpha}$, where the logarithmic
slopes are related by  $s =2\alpha~-~1$. In the strong shock, high Mach
number limit, s=2 ($\alpha$=-0.50).

The low frequency spectral slope is also important for  Inverse
Compton studies of radio galaxy lobes (e.g. \citet{feifor, kanfor}),
where the synchrotron radiating relativistic electrons upscatter
the cosmic microwave background photons to X-ray energies.
In $\mu$G fields, the keV  X-rays  are produced by electrons
radiating at $\approx 10$~MHz, far below the observed range.
Thus, an accurate extrapolation of the low frequency synchrotron
spectrum (energy distribution) is important, e.g., in combining
X-ray and synchrotron measurements to derive the magnetic field
strength \citep{grindhar}.  The low frequency slope assumes even
greater importance for recent calculations of beaming factors
in relativistic jets, \citep{HandK}, where a change in slope of
0.2 leads to an order of magnitude change in beaming factor.

\subsection{Source selection} 

We present observations of three tailed sources (B2 1116+28, B2
  1243+26 - itself two sources, and B2 1553+24), one ``fat double"
  (3C386) and one ``classical double" (3C98). The tailed sources are
  common at lower luminosities and are structurally classified as FRIs
  \citep{fan74}.  For tailed sources, most spectral observations in
  the literature are confused by the blending of emission from the
  jets with that of their steeper spectrum surrounding sheaths
  \citep{kat97,giz03}.  The tailed sources studied here were chosen
  from the work of \citet{mor+97a} where the change in polarization
  direction from the inner jets (longitudinal field) to the outer
  sheaths (transverse) showed that these components could be isolated
  at reasonable resolutions.  3C98 was chosen as one representative of
  FRIIs because of its mixture of both high and low surface brightness
  material and preliminary indications of Inverse Compton X-rays
  (N. Isobe, private communication). 3C386, a ``fat" or ``relaxed" double, provides an
  example where the absence of jets and hot spots  probably
  indicates no current relativistic particle acceleration. Although
  the images of 1553+24 are shown here, the spectral results were not
  reliable due to imaging artifacts, and are not included here.

\section{Observations and Images}

All sources were observed with  multiple configurations and frequencies of 
the Very
Large Array (VLA\footnote{The Very Large Array is a facility of the
National Radio Astronomy Observatory, operated by Associated
Universities, Inc., under contract with the National Science
Foundation.}) between November 1998 and August 2001. Table
\ref{jetobstable} provides the observational details.  Our data were
supplemented by archived data at the National Radio Astronomy
Observatory (see Table \ref{addvlatable}).  For the new observations of
sources  1116+28,
1243+26 and 1553+24, the bandwidth was 50~MHz at all wavelengths.
 For 3C98 and 3C386, the bandwidth was 3.13 MHz at 90 cm, 25 MHz at
20 cm, and 50 MHz at 6 cm. The bandwidths were consistent for all
configurations. 
 While the nominal
bandwidth at 90 cm was 3.13 MHz, the effective bandwidth is reduced
through the editing process.
 Except as noted, all observations were
taken using two IFs.  For 1243+26, one IF at 20 cm in A array was
dropped due to poor quality.  For 3C98 and 3C386, the two IFs at 20cm
provided maps of different qualities, so only one IF was used for
each. At 90cm, only one IF was used for mapping. When more than one object was accessible at
a given time, cycling snapshot observations between sources were employed to
maximize hour angle and related uv-plane coverage.

\begin{table}[]
\caption{Observations Summary}
\label{jetobstable}
\begin{center}
\begin{tabular}{l c c c c}
\hline
\hline
B2 1116+28 & Map Freq. (MHz) & 1464.9 & 4860.1 & 8460.1 \\
Dates  & Configuration & 20 cm & 6 cm & 3.6 cm \\
\hline
Sep 1999 & A & 0.96 & - & - \\
Feb 2000  & B & 0.52 & 0.75 & - \\
May 1984; Apr 1992; Nov/Dec 1998  & C & 0.47 & 0.62 & 1.52 \\
Aug 1992; Apr 1999 & D & - & 0.6 & 0.7 \\
\hline
\hline
B2 1243+26 & Map Freq. (MHz) & 1410.9 & 4860.1 & 8460.1 \\
Dates & Configuration & 20 cm & 6 cm & 3.6 cm \\
\hline
Sep 1999; Jun 1990 & A & 0.99 & 1.29$^a$ & - \\
Feb 2000;, Aug 1990 & B & 0.48 & 1.03 & - \\
Dec 1998 / Jan 1999  & C & 0.35 & 0.14 & 2.67 \\
Apr 1999; Dec 1989 \& Apr 1999; Apr 1999 & D & 0.20 & 0.99 & 0.5 \\
\hline
\hline
B2 1553+24 & Map Freq. (MHz) & 1425 & 4860.1 & 8460.1 \\
Dates & Configuration & 20 cm & 6 cm & 3.6 cm \\
\hline
Sep 1999 & A & 0.98 & - & - \\
Feb 2000   & B & 0.51 & 1.59 & - \\
Multiple$^b$ & C & 1.41 & 0.74 & 3.47 \\
Apr 1999  & D & - & 0.38 & 0.5 \\
\hline
\hline
3C98 & Map Freq. (MHz) & 327.45 & 1564.9 & 4860.1 \\
Dates & Configuration & 90 cm & 20 cm & 6 cm \\
\hline
Jan 2001 & A & 3.93 & - & - \\
Mar 2001 & B & 2.77 & 1.43 & - \\
Aug 2001  & C & 0.49 & 0.41 & 1.22 \\
Sep 1990  & D & - & 0.33 & 0.57 \\
\hline
\hline
3C386 & Map Freq. (MHz) & 327.45 & 1400 & 4860.1 \\
Dates  & Configuration & 90 cm & 20 cm & 6 cm \\
\hline
Mar 2001 & B & 2.17 & - & - \\
Aug 2001  & C & 1.10 & 0.32 & - \\
Sep 2000  & D & - & 0.24 & 1.79 \\
\hline
\end{tabular}

\end{center}
\vspace*{.6cm}
\noindent
 The entries under each wavelength are the integration times
in hours.

 Single dates apply to all entries on the row.
Multiple dates refer to the observations at different
frequencies (separated by semicolons) in column order. 

$^a$ BnA hybrid array.

$^b$ All three frequencies were observed in  Dec 1998 and Jan 1999; additional
observations were taken in Sept 1998 (20cm),  Jul 1985 (6cm), and  May
2000 (3.6cm).

\end{table}

All maps were created using standard calibration, mapping and
self-calibration techniques in common use at the Very Large 
Array.  However, given the significance of the spectral results,
we include a more detailed discussion of the flux calibration
below.  Data were obtained at 74 MHz (4m) and 330 MHz (90cm) in
spectral line mode, which facilitated removal of radio frequency
interference.  Afterward the data were averaged in frequency, with
effective bandwidths up to 30\% smaller than the nominal bandwidths
listed above.

In producing maps for all spectral index determinations, the
  uv coverage was matched, as well as the 
restoring convolution beams.
The ``clean'' algorithm (IMAGR in AIPS) was used for all maps except
for the 90cm maps of 3C98 and 3C386.  In those cases, the maximum
entropy routine VTESS was used because it
 reduced the amount of 'striping' artifacts in the final maps.

  Table \ref{b2table} contains the basic parameters for these
sources. Throughout this paper we have assumed a flat cosmology and a
Hubble constant of $H_{0}$ = 75 km/s/Mpc.  \\

\begin{table}[t]
\caption{Summary of Source Properties}
\label{b2table}
\begin{center}
\begin{tabular}{c c c c c c}
\hline 
Parameter & 1116+28 & 1243+26 & 1553+24 & 3C98 & 3C386\\
\hline 
RA (2000) & 11 18 59.392 & 12 46 21.8210 & 15 56 3.929 &  3 57 54.444 & 18 38 26.285 \\
Dec (2000) & 27 54 7.8 & 26 27 16.4 & 24 26 53.236 & 10 26 2.52 &  17 11 49.392 \\
z & 0.0672$^a$ & 0.0872$^b$ & 0.0426$^c$ & 0.0305$^d$ & 0.0170$^d$ \\
Major Axis (kpc) & 358 & 711 & 248 & 236 & 99 \\
 Integrated $I_{90cm}$ (Jy) & - &- &- & 28.9 & 16.4 \\
Integrated $I_{20cm}$  (Jy) & 0.44 & 0.32 & 0.19 & 11.1 & 6.3 \\
Integrated $I_{6cm}$ (Jy) & 0.19 & 0.15 & 0.13 & 3.45 & 2.55 \\
Integrated $I_{3.6cm}$ (Jy) & 0.11 & 0.11 & 0.11 & - & - \\
 Integrated  $\alpha^{90}_{20}$ &- & - &-  & -0.61 & -0.66  \\
Integrated  $\alpha^{20}_{6}$ &  -0.69 & -0.62 & -0.31 & -1.03 & -0.73 \\
Integrated  $\alpha^{6}_{3.6}$ & -0.92 & -0.51 & -0.22 & - & - \\
Fit Low Freq. Index & -0.54 $\pm$ 0.04  & -0.50 $\pm$0.04 & - & -0.65 $\pm$ .01 & -0.55 $\pm$ 0.02\\
Chi-squared / d.o.f. & 0.8 & 1 & - & 3.1  & 3.4 \\
\hline 
\end{tabular} 
\end{center}
\vspace*{.6cm}
\noindent
a. \cite{kar90};  b. \cite{bar98}; c. \cite{deV91}; d. \cite{mil99}
\end{table}

 \begin{table}[t]
\caption{Archived data used for images}
\label{addvlatable}
\begin{center}
\begin{tabular}{c c c c c}
\hline 
Source & Wavelength &  Configuration & Observer & Year \\
\hline 
1116+28 & 20 cm & C & P. Parma & 1984 \\
1116+28 & 6 cm & C & P. Parma & 1992 \\
1116+28 & 6 cm & D & P. Parma & 1992 \\
1243+26 & 6 cm & BnA & S. Capetti & 1990 \\
1243+26 & 6 cm & B & S. Capetti & 1990 \\
1243+26 & 6 cm & D & S. Capetti & 1989 \\
1553+24 & 6 cm & C & P. Parma & 1985 \\
1553+24 & 3.6 cm & C & R. Laing & 2000 \\
\hline 
\end{tabular}         
\end{center}

\end{table}

Images of the five sources are presented in Figures 1-10.  For the
tailed sources, we show both high and low resolution images.  Single
resolution images of 3C386 and 3C98 are shown along with color images
indicating the spectral variations within the sources.

\section{Spectral Analysis and Results}

Matched maps at three frequencies were used to create pairs of
spectral index maps (e.g., $\alpha^{3.6}_{6}$ and
$\alpha^{6}_{20}$). The pairs of indices were then sampled to
construct color-color diagrams \citep{kat93}, as shown in Figure
\ref{color1}. See Section 1.2 for the description of how color-color
diagrams are used to determine the low frequency slopes.  

 Several quality checking procedures were employed to improve the
reliability of the spectral index measurements used here. Maps were
first intensity scaled and subtracted in pairs to inspect for small
spatial misalignments; several small corrections ($\ll$ beam size) were
made.  Scatter plots of intensity at one frequency vs. intensity at
the second frequency were inspected to ensure that there was no
low-level bias on the images at a level that would contribute spectral
errors comparable to those due to the random noise.

For 3C98 and 3C386, the spectral index maps were sampled at one point per independent beam and used to construct the color-color diagrams.
Only those points with spectral errors between the low pair of frequencies
 less than 0.035 were included for 3C98 (less than 0.02 for 3C386). 
We performed a tomography analysis \citep{kat97} to assess the
possible confusion from overlapping regions of different spectral
indices.  In 3C98, there appeared to be only minor confusion (within
the errors) for the brighter regions used in the color-color diagrams.
The hot spots, for example, completely dominate their locally measured
spectral indices and fall in the same region of the color-color
diagram as the lower brightness regions.

The situation for 3C386 is more complicated, since there are gentle
spectral gradients throughout the lobes, and it is difficult to
cleanly isolate distinct spectral components (see Figure \ref{3c386plcbandb}).
Therefore, some blending of spectra is probably occurring; this will
broaden the spectral shapes and move the data towards and more 
parallel to the power law
line.  Given our fitting procedure with JP shapes, this would have the
effect of  steepening the inferred low frequency index.  However,
since the slope of the data in the color-color diagram still closely
follows the expected JP slope, it appears that any bias is within the
errors.
  
For 1116+28, 1243+26 and 1553+24, the fluxes in boxes along the jet
were used for the spectral measurements.  The use of boxes slightly
wider than the jets reduced the sensitivity to any residual
misalignments between the images.  Jet boxes were used only where any
possible contamination from overlapping steeper spectrum diffuse
emission (as assessed using tomography) would be less than the errors
due to random noise. The spectral shape results for 1553+24 appear to
be strongly influenced by imaging artifacts and are not presented
here.  For example, the spectra were dependent on the exact choice of
boxes, and for some box choices, even showed concave shapes.

\subsection{Sources from the literature}

We examined the literature for data matching the criteria for
determination of low frequency indices as described above -- high
resolution spectra with at least three points in the low frequency
power law region, matched resolution images at three frequencies
suitable for color-color analysis, and asymptotic two point spectra
approaching the nucleus.  Following is a brief description of the
sources so identified and our determination of the low frequency power
law.  The derived low frequency indices for the literature sources are
listed in Table \ref{addspix}.

For two sources, we performed a color-color analysis using published
data to determine the low frequency index.  Hercules A has recently
been mapped by \cite{giz03} at three frequencies, and the jet spectra
separated from the steeper background using a tomographic analysis.
Taking the spectral measurements from their Table 5, we found a best
fit of -0.62 $\pm$ 0.03 in color-color space (Figure \ref{color2}).  There are
several points far from the best fit line, which may indicate residual
contributions from steep spectrum material.

Hydra A was observed at 74, 330, and 1415 MHz by \cite{lane04}.  They
made no attempt to separate the emission into flatter and steeper
components.  The spectral indices were read from their Figure 5, and
plotted in color-color space in our Figure \ref{color2}.  Note that their errors,
which we reflect, correspond to 3$\sigma$, so the deviation from a
simple JP shape is quite significant.  We therefore fit only the data
with spectral indices flatter than -1.3, which yield a low-frequency
index of -0.59 $\pm$ 0.08.  The steeper data could be modeled using a
second component with a lower cutoff frequency, but would not add
useful information about the low frequency slope.

In a number of cases, spectra corresponding to the low frequency slope
had already been identified.  These include the jets isolated through
a tomography analysis: 1231+674 and 1433+553 \cite{kat+99}, 3C449
\citep{kat97} and 3C130 \citep{hard99}.  In these cases, the
two point spectra of the jets asymptote to a constant spectral index
as they approach the nucleus. (For 3C449 this behavior 
is seen between two
pairs of frequencies, so the color-color diagram is also useful.) 

 The spectrum of the inner kpc
of 3C264 was measured by \cite{lar+97} using a least-squares fit to
four frequencies, after subtracting the contribution from the flat
spectrum core. 3C264 shows a clear one-sided jet, and is noted in
Figure \ref{summary} as ``other FRI'' because of its unusual amorphous large-scale
emission \citep{lar+97}.  The radio data for M87 at 1.46, 4.89 and
14.96 GHz, are consistent with a power law, which \cite{bir91} cite as
$\alpha_{rr} \approx 0.5$ in their abstract.  We have used their data
to recalculate the spectra of knots A-I to yield a value of -0.53~$\pm$~0.04.

\begin{table}[t]
\caption{Additional low frequency spectral indices}
\label{addspix}
\begin{center}
\begin{tabular}{c c c}
\hline 
Source Name & Spectral Index & Reference\\
\hline 
1231+674 & -0.55 $\pm$0.05  & \cite{kat+99}, \cite{you+02}\\
1433+553 & -0.55 $\pm$0.05  & \cite{kat+99},\cite{you+02}\\
M87 & -0.53 $\pm$ 0.04 & \cite{bir91}\\
3C130 & -0.55 $\pm$0.05  & \cite{har99}\\
3C449 & -0.53 $\pm$0.05  & \cite{kat97}\\
Hydra A & -0.59  $\pm$ 0.08 &  \cite{lane04}\\
Hercules A & -0.62 $\pm$ 0.03 & \cite{giz03}\\
3C264 & -0.58 $\pm$ 0.03 & \cite{lara04} \\

\hline 
\hline 
\end{tabular} 
\vskip 9pt
\end{center}
Other maps and information about the optical environments of these
sources are available in the literature, and starting points for
interested readers include: 1116+28 \citep{mor+97b,par85,car97};
1243+26 \citep{mor+97b,gon93,col95,gon94}; 1553+24
\citep{mor+97b,sto87,par+03, canvin04}; 3C98
\citep{are+01,lea+97,mil+02}; 3C98 \citep{lea91,bau00,freu03}; 3C386
\citep{mil+02,sim+96}
\vspace*{.6cm}
\noindent
\end{table}

\subsection{Flux calibration}

The  low frequency index of FRIs appears to have a
characteristic value of -0.55, and not the limiting -0.50 test particle
case.  Given the scientific implications of this difference, we present a brief discussion of calibration issues to assess
its  reliability.

Following the standard flux calibration procedures at the Very Large
Array, we included observations of ``primary'' calibrators in each
observing session.  The fluxes of these calibrators are monitored
every few years and normalized to 3C295, with its value set to the
\cite{baa77} scale, for frequencies between 327~MHz and 15~GHz
\citep{vlacal}.  The uncertainties of the primary calibrator fluxes
determined by \cite{baa77} at each frequency are no more than $3-4\%$,
and our best estimate of the final uncertainties in the VLA flux scale
at each frequency are thus less than $ 5\%$.  This is a potentially
important systematic effect; if 3C295 were incorrect at the $5\%$
level at one frequency, this would lead to a spectral index error of
0.05 when the second frequency is a factor of three away.  Ultimately,
the quite difficult task of doing an absolute recalibration at better
than the few percent level would be needed to resolve this issue.

In the meantime, we have another indicator that the flux scale
accuracy is sufficient to distinguish between -0.55 and -0.50, viz.,
the observed power law spectra of Cassiopeia A.  The original
\cite{baa77} work was designed to measure the integrated spectrum of
Cas A, and they show that it is a power law, with a mean error in
scale of $\approx 2\%$ between 300~MHz and 30~GHz. In another program,
we have also been making careful spectral index maps of Cas~A using
the VLA.  We find that there is a narrow, but very well-defined range
of spectral indices across the remnant \citep{and91, and96}.
Our more recent work \citep{bow05} extends these studies to longer
wavelengths. We find that the spectra at different locations in Cas~A
are each well-described by power laws between wavelengths of 90, 20,
and 6cm, and extending to 4m in locations where absorption is not
important.  Below, we describe the procedures leading to this result,
and its significance for the accuracy of the flux scales.

We followed the standard flux calibration procedures for Cas~A, except
that an additional correction is required at 90cm because of the
saturation of the digital samplers due to the high correlated fluxes
at short baselines.  We determined the amplitude of this ``Van Vleck"
correction \citep{VV} to be $\approx 3\%$ by measuring the change
in total flux on the images when the short baseline data were removed.
Applying this correction to the 90cm data, we found that the data were
consistent within the errors with $\alpha_{4m}^{90} =
\alpha^{20}_{90}$ for the remnant's exterior, while $\alpha^{90}_{4m}
> \alpha^{20}_{90}$ (i.e. flatter at long wavelengths) for the
interior, where absorption is important \citep{kas95}. This gives
some confidence that the flux scales are correct, but is based on the
assumption, for example, that there is no absorption for the exterior
regions of the remnant.

Much tighter constraints are seen when we compare the spectral indices
between 90, 20, and 6cm.  We find local power laws (i.e.,
$\alpha^{20}_{90} \approx \alpha^6_{20}$) ranging from -0.82 to -0.73
at 240 independent locations. We find that $\delta\alpha =
\alpha^{20}_{90} - \alpha_6^{20}$ had a mean of 0.0006 with an rms
scatter per point of 0.014.  To understand the significance of this,
suppose that the flux scale at wavelength $\lambda_i$ is incorrect by
some factor $g_i \neq 1$. Then the error in spectral index between
$\lambda_i$ and $\lambda_j$, $\delta\alpha_i^j$ is given by:
$$\delta\alpha_i^j = \frac{ ln(g_i/g_j)}{ln(\lambda_i/\lambda_j)}. $$
Since we observe that $\alpha^{20}_{90} \approx \alpha^6_{20}$, this
implies that
$$ \frac{ln(g_{90}/g_{20})}{ ln(90/20)} = \frac{ln(g_{20}/g_6)}{
ln(20/6)}.$$ It is thus possible that our flux scales are all
incorrect, but that the correction factors fortuitously preserve power
laws between 90, 20 and 6cm.  It seems more likely that each of the
flux scales is correct, yielding spectral indices accurate to within
2\%, and thus enabling us to distinguish between -0.50 and -0.55.


\subsection{Other jet spectra}

A number of additional sources had some multi-frequency data available,
but were not included in our summary, due to a variety of different
problems.  The well-studied radio jet in 3C273, for example, has
matched multi-frequency measurements \citep{con93} but can not be
described in terms of a single electron population, as can be seen in
Figure \ref{273cc}. The points above the power-law line (where $\alpha^2_6 =
\alpha^6_{18}$) represent places in the spectrum with concave
curvature, i.e., flatter at higher frequencies.  This would result,
e.g., from a combination of two different power law spectra.  Multiple
components are also suggested by the radio, \citep{con93}, UV and
X-ray  measurements of 3C273 by \cite{jes+01}.

Extensive multi-frequency data are also available for the tailed radio
galaxy 3C31 \citep{lai02,lai04}, who suggested an approximate spectral
index of -0.55 for the jets.  Based on fits from 1.4~GHz to 8.4~GHz at
1.5 arcsecond resolution, the regions close to the nucleus in the
North (South) show power laws of -0.59 (-0.60) respectively (R. Laing
and A. Bridle, private communication). Further from the nucleus, the
spectra retain a power law shape, but with generally steeper spectra.
Therefore, the steepening along these jets is not simply due to
synchrotron losses from a homogeneous electron distribution, which would
show curved spectra, and so  a
unique low frequency slope cannot be derived.

The presence of multiple and distinct power laws can arise in two
ways.  First, it could represent the low energy spectral slope from
different acceleration conditions at different positions along the
jets/tails. Second, power laws over less than a decade in frequency
can result from a mixture of different magnetic fields and/or losses
within the integration region. Given the clear presence of a
spine/sheath structure \cite{laingref}, or, more generally, a
non-uniform transverse velocity profile \citep{lai02} the mixture
scenario is more likely.  Again, in this case, it is not yet possible
to isolate a unique low frequency slope.

 Other examples of ambiguous spectral shapes include 3C66B, where
 \cite{har96} describe the inner jet spectra ranging from
 $\approx$~-0.5 to -0.6, while \cite{har01} cite much steeper spectral
 indices for Knot A.  Three frequency data at a common resolution have
 not been published for this source, and would be especially useful
 because of the infrared and X-ray emission from the jet
 \citep{tan00,har01}.  The BL Lac object PKS 0521-365 has
 three-frequency radio data with a spectral index of
 $\approx$-0.6 \citep{bir02}, but the authors suggest that this is
 contaminated by surrounding steeper spectrum emission.

Of the three tailed sources studied here, only 1553+24 has a detected
optical jet \citep{par+03}. For the brightest knot,
$\alpha^{optical}_{radio} \approx$ -0.67.  Radio-optical spectral
indices are generally not useful for examining the low-energy slope of
the electron distribution, since they can be easily dominated by
radiative losses and possible later acceleration processes
\citep{par+03}.  These problems are even more severe for electrons
radiating in X-rays. M87 illustrates the confusion from
losses and reacceleration.  Its average $\alpha_{radio}$ is -0.53,
while the $\alpha^{optical}_{radio}$ average is -0.67, and using
optical measurements alone, $\alpha_{optical}$ = -0.9
\cite[]{per+01}. This steepening is not simply related to radiative
losses, because there is a poor correlation between $\alpha_{optical}$
and $\alpha_{radio}^{optical}$, and $\alpha_{optical}$ flattens and
steepens all along the jet. \cite{per+01} believe that the variations
are evidence of ongoing particle acceleration at the brightest points
in the jet.

A $\approx$1~kpc optical jet has also been found in 3C264 (NGC3862,
\cite{bau97,lar+97}).  $\alpha_{radio}^{optical}$ varies only slightly
along the jet, although the overall spectrum steepens dramatically
from $\alpha_{radio} \approx -0.46$ to $\alpha_{optical} \approx -1.34$
\citep{lar99}.  They suggest that local reacceleration of relativistic
electrons along the jet may be responsible.  No low frequency index
can currently be identified.

\subsection{Summary of low frequency spectra}
The low frequency spectra currently available are summarized in Figure
\ref{summary} with our best estimates of their errors. Almost all of
the FRIs are consistent with a low frequency value of -0.55.  This is
seen in sharp distinction to the distribution of integrated spectral
indices for sources from the B2.3 catalog \citep{kul85a}, as shown in
Figure \ref{alphahist}. The latter have a median value of $\approx
-0.93$, which varies slightly with flux density \citep{kul85b} and a
very broad spread.  These integrated spectral indices depend on the
long-term history of particle injection, reacceleration and the losses
in magnetic fields of varying strength, as well as the current distribution
of magnetic fields strengths, which broadens the spectral shape. The narrow range of low
frequency indices suggests a canonical acceleration process, at least
for the existing small sample of  FRIs.

It is important that these results be confirmed using a larger sample.
Partial data exist for a number of sources, and matched resolution
observations at other frequencies are practical.  Such sources include
well-known tailed radio galaxies such as NGC1265 \citep{odea86}, 3C465 \citep{eil02},
and 3C129 \citep{tay01}.  In addition, there are a  number of other
sources accessible to current instruments that could be selected from
published works  (e.g., J0448-2025 in Abell 514, \cite{gov01}, and NGC326 at 1'' resolution, \cite{mur01}).
We note, however, that great care must be done in the analysis to 
remove subtle artifacts and especially 
 to isolate overlapping spectral components, as
discussed earlier.  In addition, even with careful work, some of these
may not yield a single well-defined low frequency index, such as
3C31 and 3C66B, as discussed above.


\subsection{A note regarding FRIIs}
The results on 3C98, the sole FRII in our sample, show the steepest
 apparent low frequency synchrotron  slope of -0.65 $\pm$ 0.01 (energy
index of -2.3 $\pm$  0.02). Its small range in observed spectral indices, far from
the power law line, leaves open the possibility of a more complicated
spectral shape than modeled here.  This can only be determined with lower frequency
measurements.  If we assume a simple spectral shape, we find that each hot
spot shows two-point spectra steeper than the inferred low frequency index. 
At the same time, the spectral {\it curvatures} of the hot spots are consistent
with a single electron population for both them and the lobes. This poses a dilemma
for the northern hot spot, which has steeper two point spectra than the lobes (at
the lower end of the color-color distribution).  If, as commonly assumed, the lobes
are populated by electrons that have passed through the hot spot region, then
adiabatic and radiative losses should steepen the spectra in the lobes.  In addition,
any reduction in magnetic field strength into the lobes implies that we are
observing higher energy electrons at a fixed frequency, which should show steeper
spectra.  It is thus not clear how the northern hot spot and lobe electron populations
are related.

The limited data on other hot spot spectral shapes also presents
a confusing picture. The spectra of Cygnus A's hot spots
 show curvature at the low end, so there is no clear low energy power
 law slope \citep{car91,kat93}.  As another example, 3C295's hot spots  have
 power-law spectral indices of -1.0 and -0.89 \citep{tay92}. However,
 it is not possible to connect these with the rest of the source
 spectra because of the combination of free-free absorption and
 radiative losses.  Unpublished work by \cite{gru01}, using data from
 \cite{blu00} yields inferred low frequency indices of -0.85 for
 3C356, no suitable fit for steep spectrum hot spots in 3C171, and an
 index of -0.5 for one hot spot in 3C172. Hot spot spectra probably
 result from both acceleration and high losses \citep{mei+89}.  In
 addition, a range of spectral indices is probably produced through
 acceleration in the 'shock web' discussed by \cite{tre01}, instead of
 a simple jet terminal shock.

Other multi-frequency data on jets in FRII sources, which are less
 prominent than the FRI jets, are rare and difficult to disentangle
 from the surrounding emission.  In the studies of 3C401 by
 \cite{tre+01}, the south jet maintains a constant two-frequency
 spectral index of -0.55, while the north jet varies from -0.4 to
 -0.9. The south jet of 3C438 also varies from -0.5 to -0.8. From the
 work of \cite{sch+01}, the jet of 3C326 varies from -0.5 to -1.5.

\section{Discussion}

As introduced earlier, the need for (re-) acceleration on kpc scales
is well-established from observations of X-ray synchrotron jets.
The key result  presented here is that the low frequency
indices of FRI jets appear to have a fairly narrow distribution around
a value of -0.55.  In particular, the average value appears to be steeper than
the -0.50 test particle strong shock limit for diffusive shock
acceleration in non-relativistic shocks.  The challenge, then, is to
find an acceleration process that is robust to differences in physical
conditions in different sources and that yields synchrotron spectra 
(energy index) slopes  of $\approx$-0.55(-2.1).

Non-relativistic diffusive shock acceleration alone, with low Mach
numbers  to generate the steeper spectral indices, is not an
attractive option.  To produce spectra with $\alpha_{low}=-0.55 \pm
0.05$ requires a distribution that can extend to high Mach numbers,
but must cut off sharply below 4.5 (see Figure
\ref{alphamach}).  There is no obvious reason why this cutoff
should exist.

Ultra-relativistic shocks provide one mechanism for producing steeper
spectra from diffusive shock acceleration.  \cite{ach01} describe
simulations and semi-analytic calculations that show a `nearly
universal' slope to the momentum distribution of -2.2 to -2.3,
equivalent to spectral indices of -0.6 to -0.65.  For the
trans-relativistic case explored by the Monte Carlo calculations of
\cite{ell02}, the equivalent spectral index is $\approx -0.56$, as
long as the scattering is sufficiently 'fine'.  Although this looks
very promising, these results apply only to the case of flows parallel
to the shock normal.  Oblique shocks produce spectra that asymptote to
this value, but can be much steeper depending on the obliquity angle
and the scattering parameters \citep{ell04}.

Modification of the shocks by the back pressure of accelerated cosmic
rays provides another mechanism for steepening the low-energy spectra
\citep{ell91, mal01, ell02}.  The resulting spectra are concave,
although the curvature is small enough that at low energies they mimic
a power-law over ranges of order 10 in energy. Outside of the
acceleration region, radiative losses would become important.  The
small curvature at the low frequency end would be hard to detect in
the color-color diagram, and we would fit the data with standard JP
spectra with low-frequency power laws. Subshock Mach numbers are
typically around 3, yielding synchrotron slopes of $\approx~ $-0.7,
steeper than observed.  One possibility deserving further exploration
is the 'self-regulating' mechanism discussed by \cite{gie00}, which
operates through the amount of 'injection' into the relativistic
particle population.  This injection is very sensitive to the high
energy tail of the thermal distribution of electrons, but the more
particles removed from the tail (strengthening the modification), the
lower the temperature of the thermal distribution (lowering the
injection).  Of course, mechanisms other than diffusive shock
acceleration might be the important ones, as mentioned in the
Introduction.  A discussion of these is beyond the scope of this
paper.

Assuming that there is a common low frequency index of -0.55 for FRI
radio galaxies, where might it arise?  One possibility is that it
represents the initial acceleration of particles in the jet as they
emerge from the nucleus.  This runs into serious difficulties for
Hydra~A, for example.  On pc scales its jet has an integrated
spectrum that peaks around 5~GHz, probably because of free-free
absorption at low frequencies \citep{tay96}. The optically thin part
of the spectrum has a slope of $\approx -0.7$.  A power law this steep
is inconsistent with the flattest points observed at arcsec scales
(Figure \ref{color2}), and so particle reacceleration is necessary.  This situation
is exacerbated by the much weaker magnetic fields on kpc, as opposed
to pc scales, as inferred from minimum energy calculations.  Lower
magnetic fields require higher energy relativistic particles to
produce the observed GHz emission, so if anything, the kpc scale
emission should be much steeper.  Other support for {\it in situ}
acceleration comes from energetic or dynamical arguments \citep{eil96,
sib98}. Thus, reacceleration on kpc scales
seems unavoidable.

Spectra on pc scales for the other sources presented here are not
available.  However, Compact Symmetric Objects like 2352+495,
which are currently thought to be young precursors of larger objects
(FRIs?, FRIIs?) also illustrate the reacceleration challenge.  Its
lobes have spectral indices of $\approx$-1.0 in the 5-10 GHz range
\citep{rea96}, under conditions of strong minimum energy fields
($\approx$5mG).  Nearby BL Lac objects such as Markarian 501 suggest
that steep spectra $<-1$ are also found in pc scale jets, at least
outside of the brightness enhancements.

One possible site for particle reacceleration for FRIs is where the
jets flare and dramatically brighten away from the nucleus. The
pre-brightening regions were historically referred  to as ``gaps''
\citep{miley80}; they are clearly visible, e.g., in 3C449 (see Figure
1 in \cite{kat97}).  Measurements at higher resolution and frequency
show that these gap regions actually contain faint emission (e.g.,
again for 3C449, see Figure 6 in \citet{fer99}). The flaring and brightening have been long suggested to
arise from entrainment of external material coupled with the rapid
deceleration of the jet, perhaps from relativistic speeds
\citep{dey93,bic95,bow96}.  In their re-analysis of 3C31 \citet{lai04} conclude that
adiabatic models of the deceleration produce too steep of a decline in
brightness along the jets, and that the distributed injection of
relativistic particles would provide a remedy.  The presence of
decelerating relativistic jets and the need for non-adiabatic inputs
is also inferred for two other low-luminosity jets, B2 0326+39 and B2
1553+24 (one of the sources studied here) by \cite{canvin04}.

Two future investigations could test whether the relativistic electron
 populations are regulated by particle acceleration in regions where
 entrainment, deceleration, and flaring of FRI jets occurs.  From an
 observational standpoint, the low frequency index of emission in the
 fainter inner jet (gap) regions needs to be measured.  If it were
 steeper than the canonical -0.55 that we have found, this would
 provide strong support for particle (re)acceleration. Theoretically,
 we need to know whether the population of shocks and turbulence
 expected in the flaring region would statistically result in a narrow
 range of indices around -0.55.
\cite{man99} investigate the effects of acceleration in the turbulent
eddies and shocks of jets, both for second-order Fermi and shock-drift
processes.  They conclude that the emergent energy distributions
closely reflect the initial ones, so that the special -0.55 value
would still need to be explained. Models involving shear, as opposed to shocks, have also been explored  \citep{jok90,sta02,rig04} as 
well as models producing acceleration through magnetic reconnection \citep{les98}.

Since FRII jets do not undergo the rapid flaring seen in FRIs (which
is why they can produce compact hot spots far downstream), the above
type of particle acceleration would not be expected.  Acceleration in
the relativistic portions of the jets \citep{sta04} and/or in the hot
spot regions \citep{man02,bru03} would then dominate, and presumably
lead to different spectral behaviors.  Particle acceleration in these
locations is necessary, in any case, to explain the very short lived
emission seen at optical and X-ray wavelengths.

\section{Conclusions}

A canonical low frequency spectral slope of $\approx$-0.55 is found in the
small existing sample of  FRI
radio galaxies with suitable data.
 If confirmed in larger samples, this would indicate
 a common acceleration mechanism for
the relativistic particles that likely occurs outside of the nuclear
regions.  One possible site for such acceleration is the brightening
and flaring region on kpc scales, where entrainment may be
decelerating the jets.  The specific acceleration mechanism cannot yet
be identified.  The behavior of FRIs appears to be different than the
particle acceleration found in FRII jets and hot spots, although more
work in this area is needed.

\acknowledgements
This work was supported in part by National Science Foundation grant AST 03-07600 to the University of Minnesota. Basis research in radio astronomy at the Naval Research Laboratory is supported by the Office of Naval Research. We gratefully acknowledge the assistance of VLA staff, discussions with Tom Jones and Diana Worrall on scientific issues, and Gabriele Giovannini for making earlier data available.




\clearpage
\newpage
\begin{figure}[t]
\epsscale{.55}
\plotone{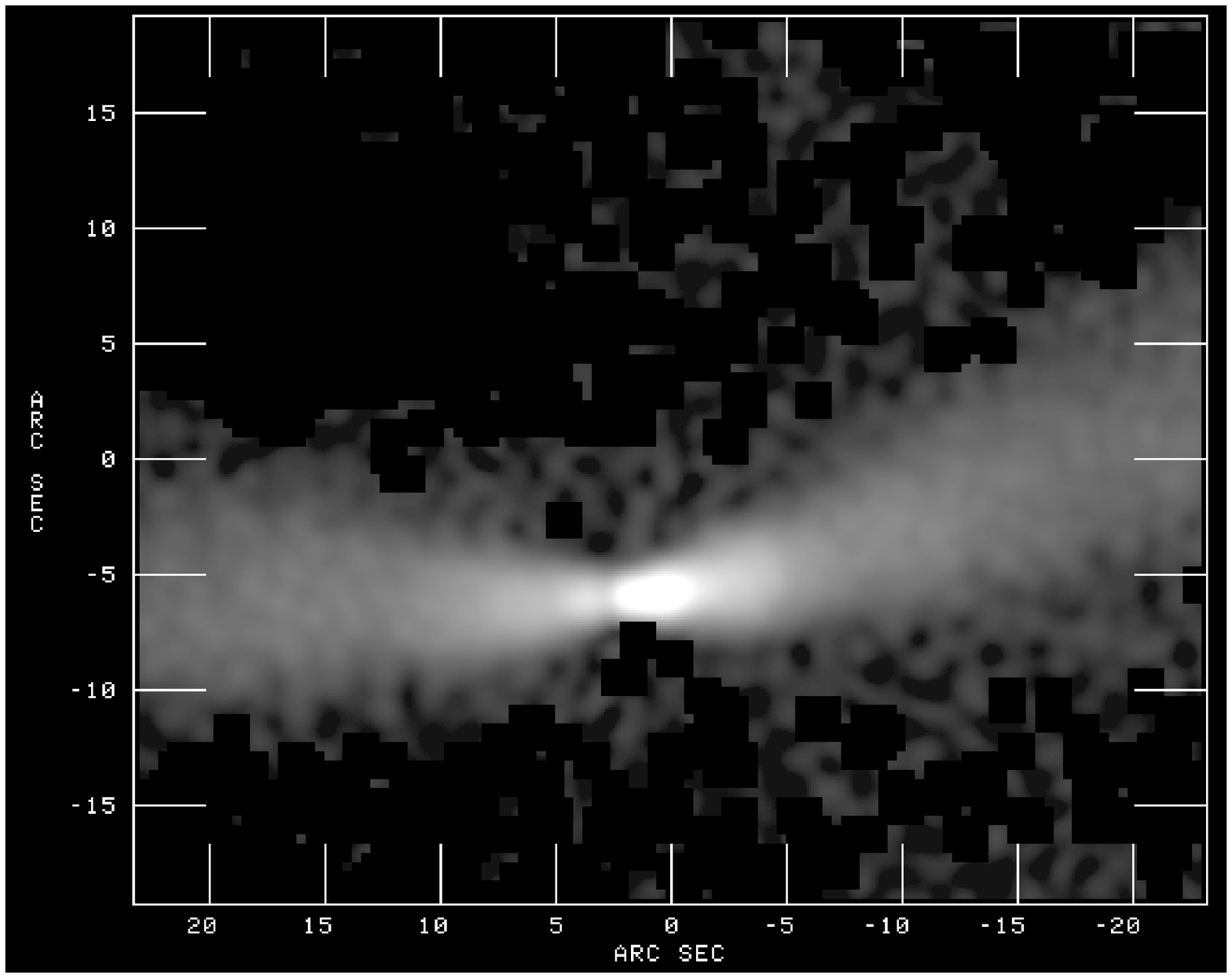}
\vspace{13pt}
\caption{B2 1116+28:  1.2'' resolution  image of inner 40  arcsec at 20cm, showing rapid flaring of the jets. Peak (rms) flux densities are
150 mJy (0.1 mJy)/beam.}
\label{1116figa}
\end{figure}

\begin{figure}[t]
\epsscale{.55}
\plotone{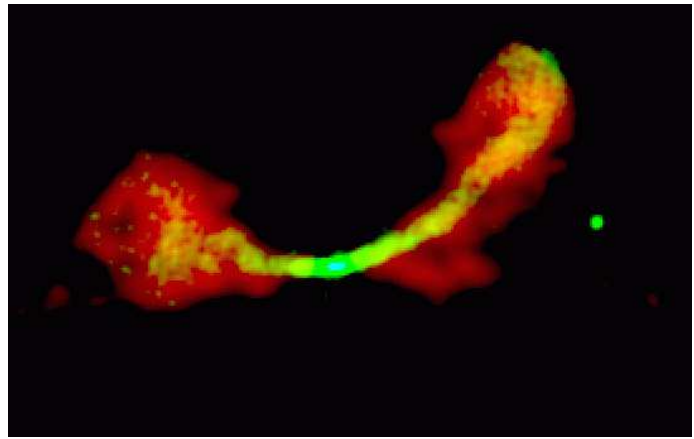}
\caption{B2 1116+28: Spectral color image of entire source. Green is
a  4'' resolution  6cm image  superposed on a red  10'' resolution
image of the difference  image (S$_{20}$ - 2$\times$S$_6$).}
\label{1116figb}
\end{figure}

\clearpage
\newpage
\begin{figure}[t]
\epsscale{.3}
\plotone{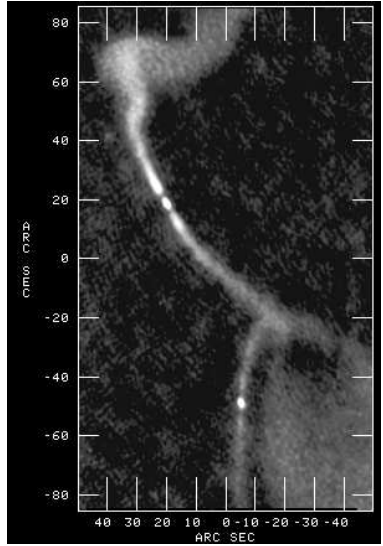}
\caption{B2 1243+26: 20cm image of inner portion at 1.3'' resolution
showing the two compact cores and ``gaps'' at the begining of the
four jets. Peak (rms) flux densities are 4.6 mJy (0.05 mJy)/beam.}
\label{1243lcxbanda}
\end{figure}

\begin{figure}[t]
\epsscale{0.3}
\plotone{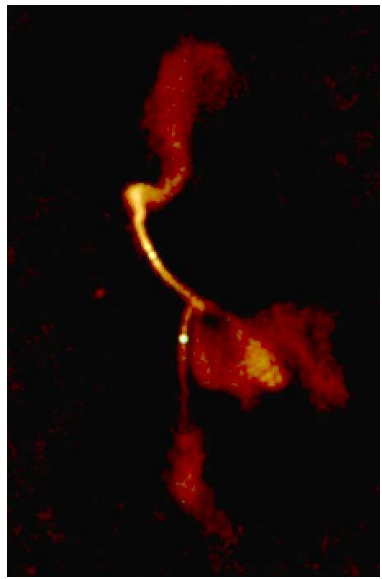}
\caption{B2 1243+26: 20cm at 4'' resolution with colors indicating
the spectral index to 6cm;  yellow (red) corresponds to -0.5 (-1.0).}
\label{1243lcxbandb}
\end{figure}

\clearpage
\newpage
\begin{figure}[t]
\epsscale{0.55}
\plotone{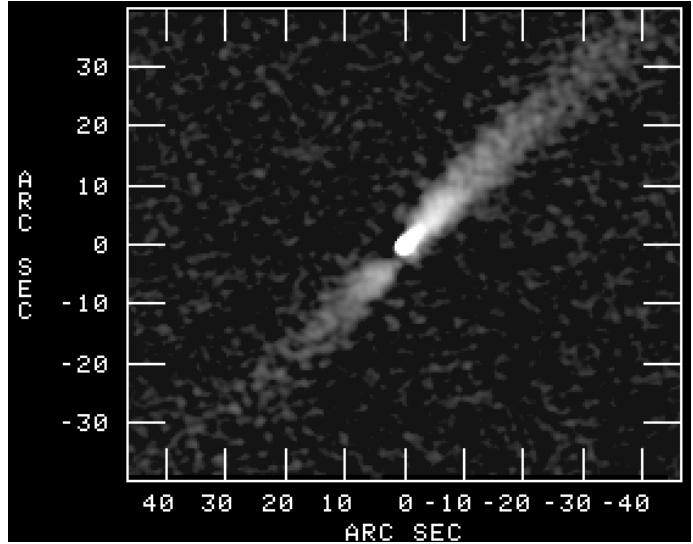}
\caption {B2 1553+24: Inner portion at 20cm, 1.4'' resolution. Peak
(rms) flux densities are 45mJy (0.045 mJy)/beam.}
\label{1553lcxband15a}
\end{figure}

\begin{figure}[t]
\epsscale{0.55}
\plotone{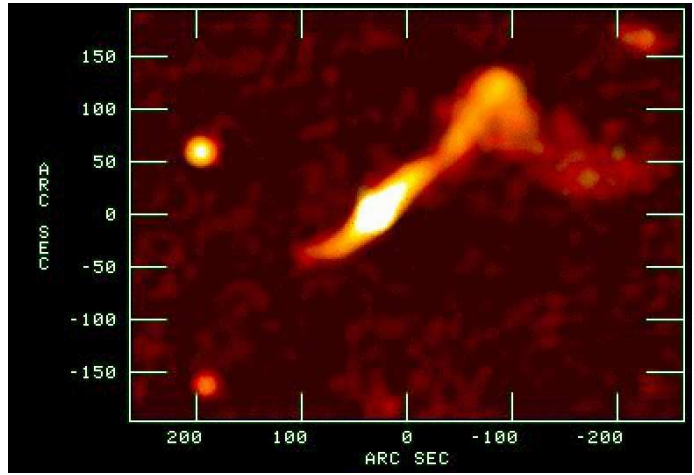}
\caption{B2 1553+24: 20cm image at 15'' resolution with
color-coded spectral indices to 6cm;  white near the core (red) 
corresponds to -0.57 (-0.78). Image has been burned out to show the diffuse emission not visible in other published maps (see \citet{canvin04} citing \citet{con98})}.
\label{1553lcxband15b}
\end{figure}

\clearpage
\begin{figure}
\begin{center}
\epsscale{0.35}
\plotone{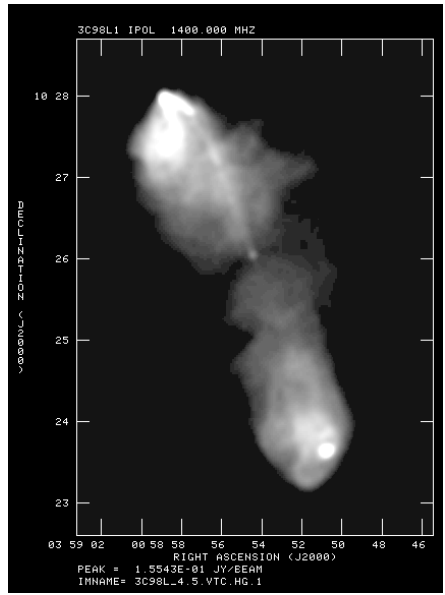}
\end{center}
\caption{3C98: image at  20 cm, 4.5'' resolution.
Peak (rms) flux densities are 155 mJy (0.34 mJy)/beam.}
\label{3c98plcbanda}
\end{figure}

\begin{figure}
\begin{center}
\epsscale{0.35}
\plotone{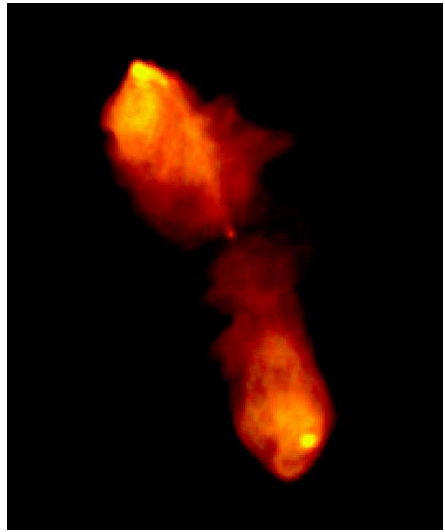}
\end{center}
\caption{3C98:  20 cm, 4.5'' resolution color-coded by spectrum
to 6cm;  yellow (red) corresponds to -0.7 (-1.1).}
\label{3c98plcbandb}
\end{figure}

\clearpage
\newpage
\begin{figure}
\begin{center}
\epsscale{0.35}
\plotone{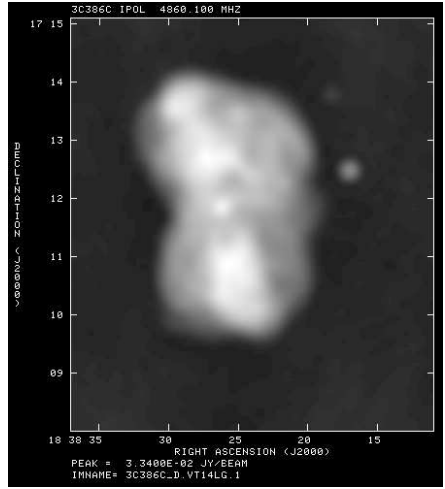}
\end{center}
\caption{3C386:  6 cm image at 14'' resolution. Peak (rms) flux densities are 33 mJy (0.085 mJy)/beam.}
\label{3c386plcbanda}
\end{figure}

\begin{figure}
\begin{center}
\epsscale{0.35}
\plotone{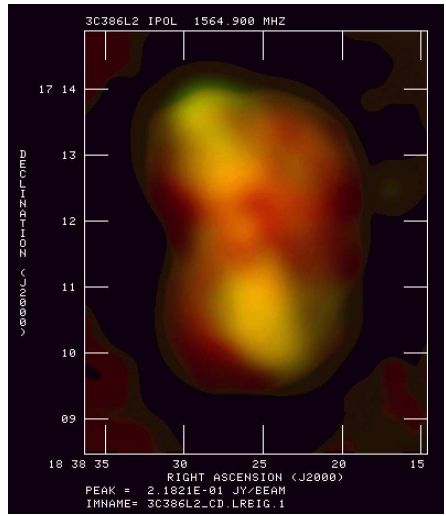}
\end{center}
\caption{3C386:  Spectral index coded intensity, where the visible
indices range from approximately -0.7 (yellow-green) to -0.9 (red).}
\label{3c386plcbandb}
\end{figure}

\clearpage
\newpage
\begin{figure}[t]
\begin{center}
\epsscale{0.45}
\plotone{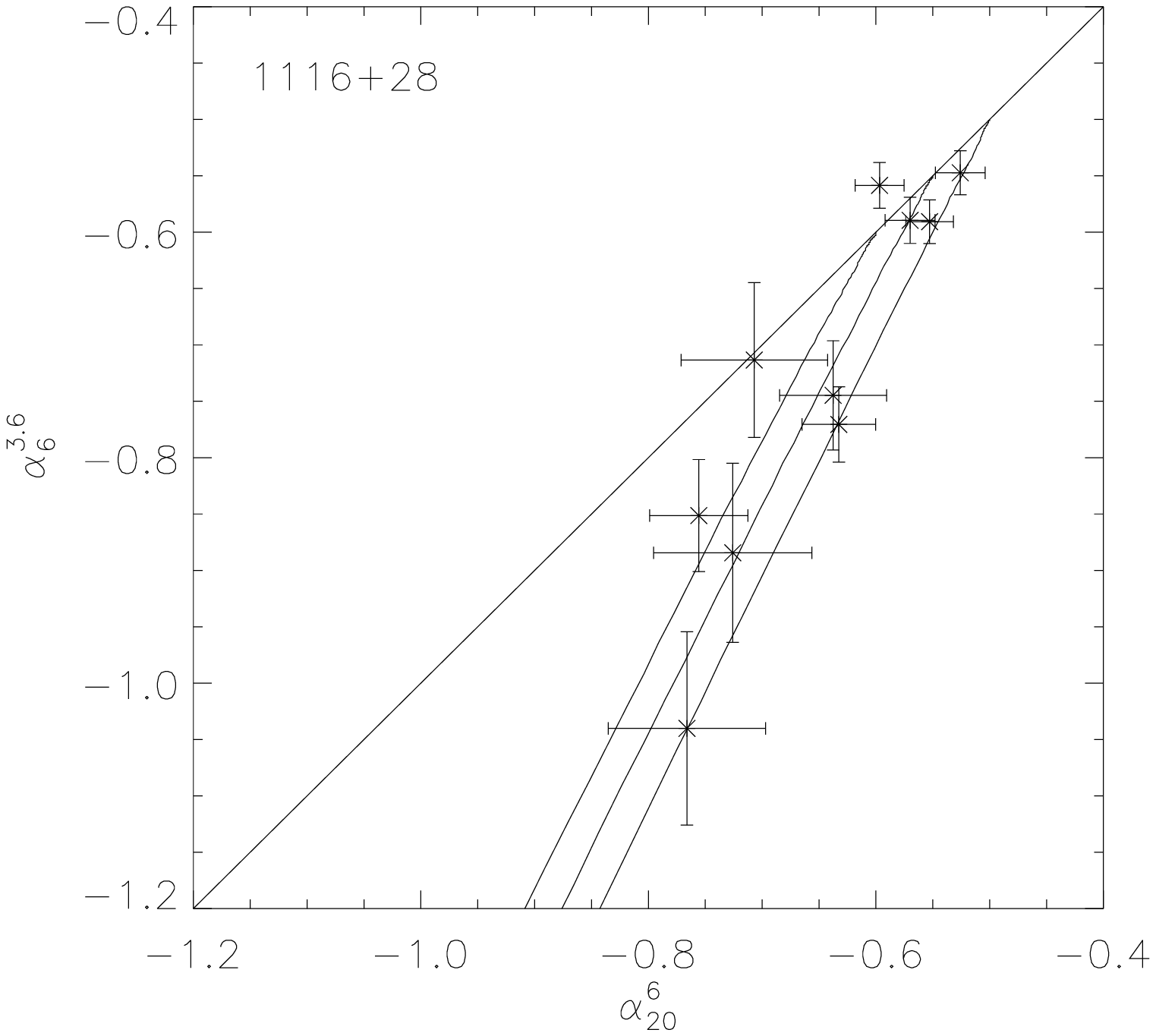}

\vspace{9pt}
\plotone{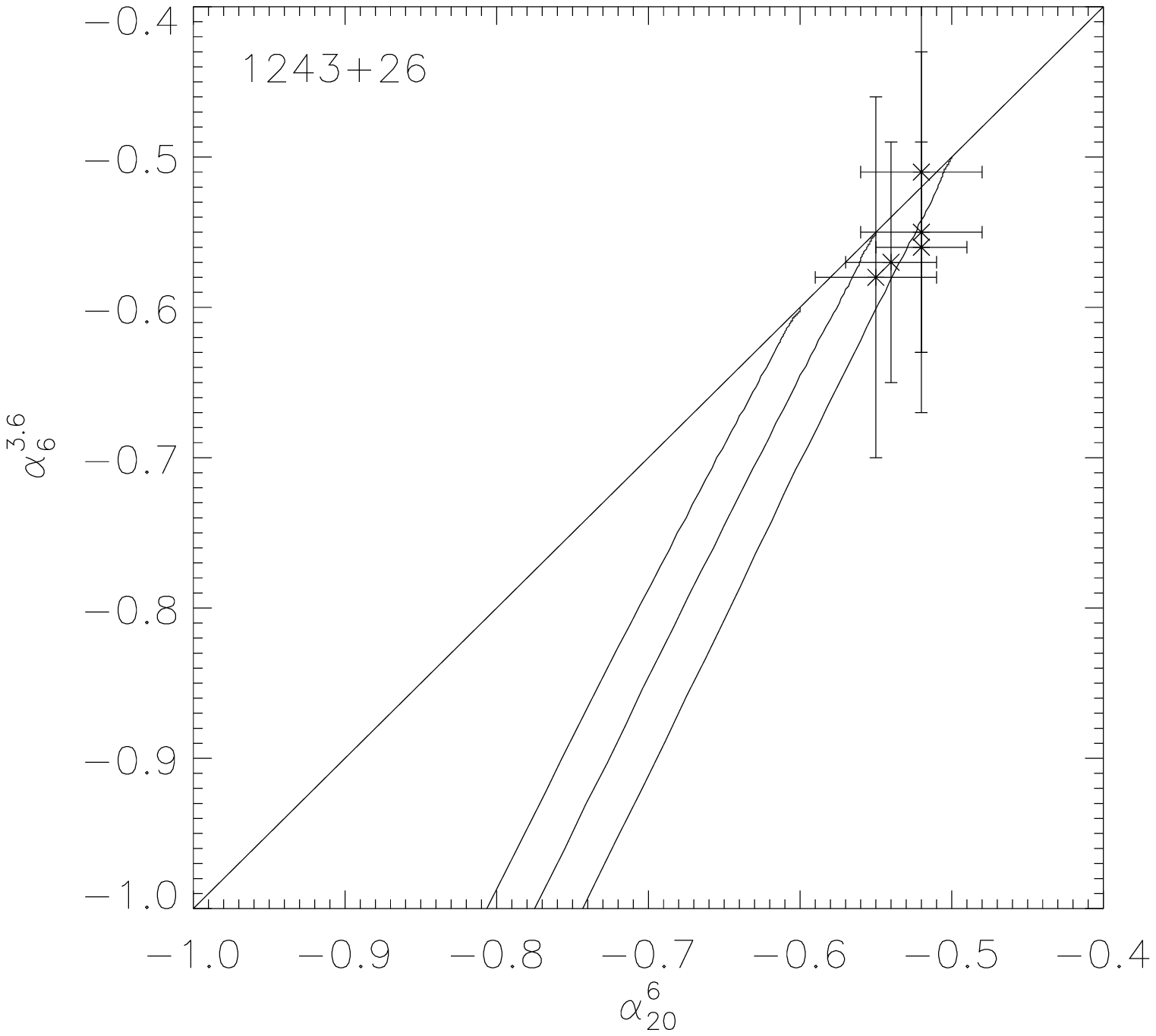}
\vspace{9pt}
\plotone{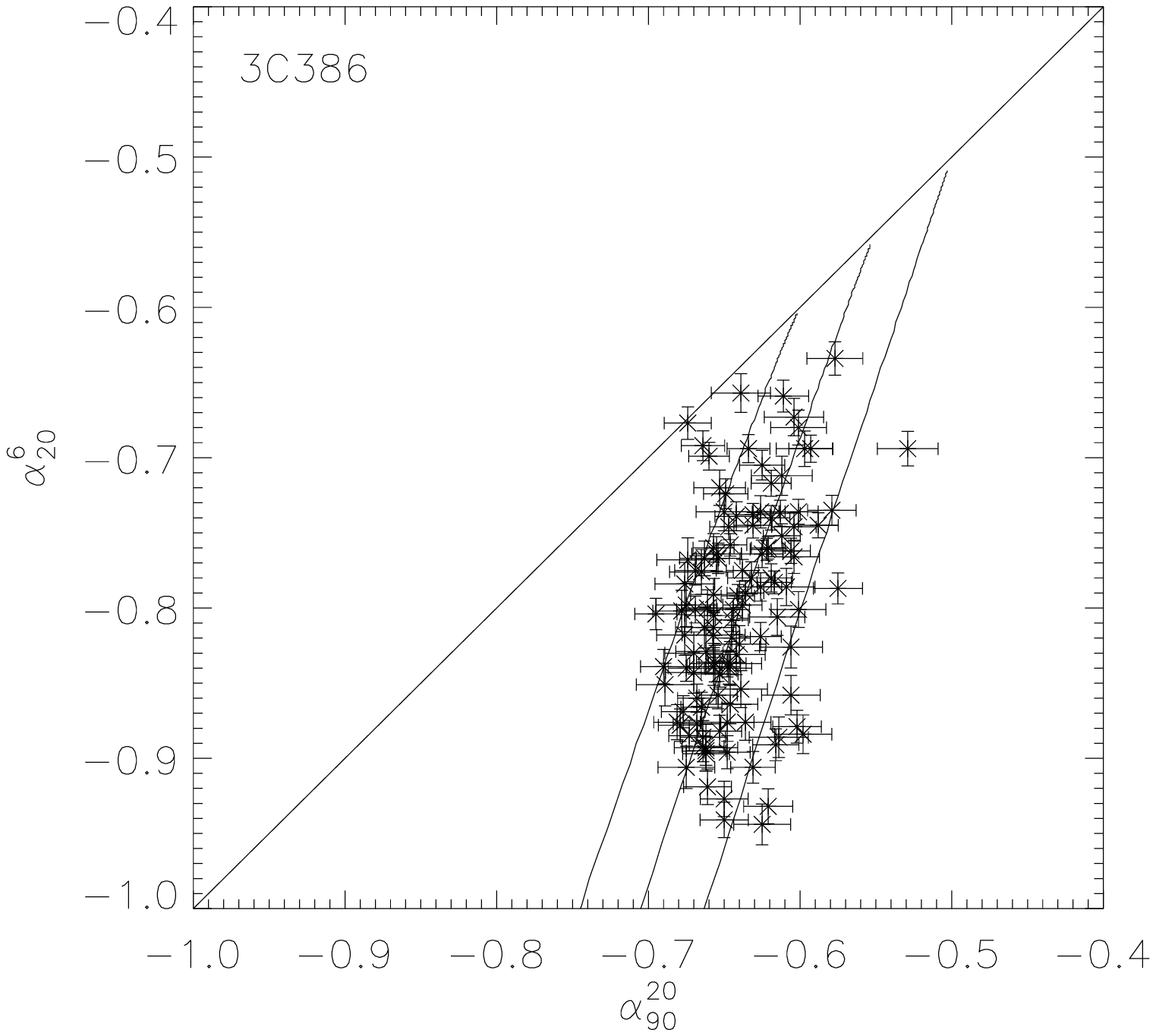}
\vspace{9pt}
\plotone{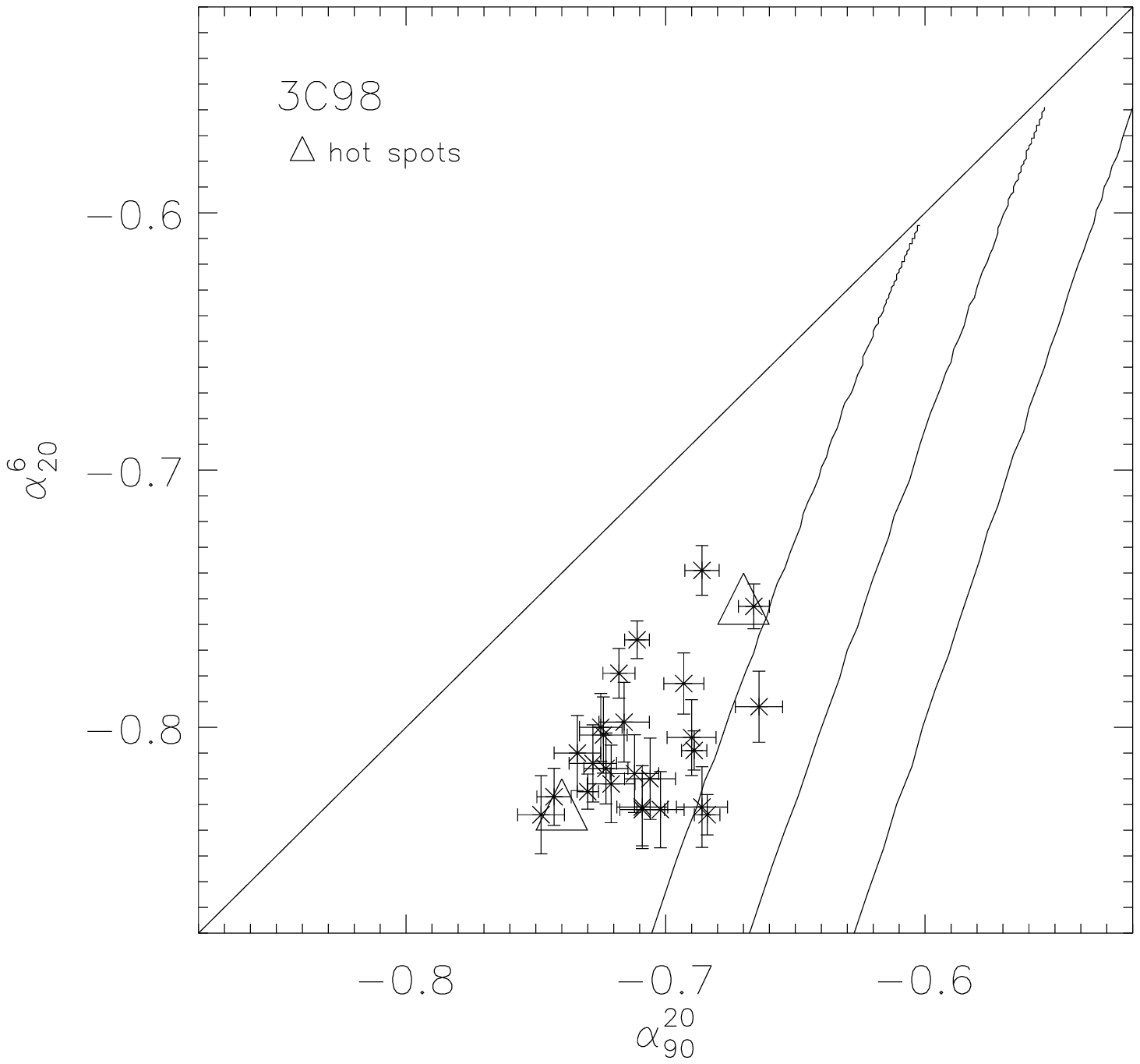}

\end{center}
\caption{Color-color diagrams (resolution):  1116+28 (4''); 
1243+26: (4''); 3C98 (4.5''); 3C386 (14'' resolution).  The diagonal
lines connecting the lower left and upper right corners are the
``power law'' lines, i.e., the locus of power law spectra with
different slopes.  The three other lines intersecting the data 
represent JP spectra, within initial slopes of -0.60, -0.55 and
-0.50 from left to right.}
\label{color1}
\end{figure}

\begin{figure}[t]
\begin{center}
\epsscale{0.4}
\plotone{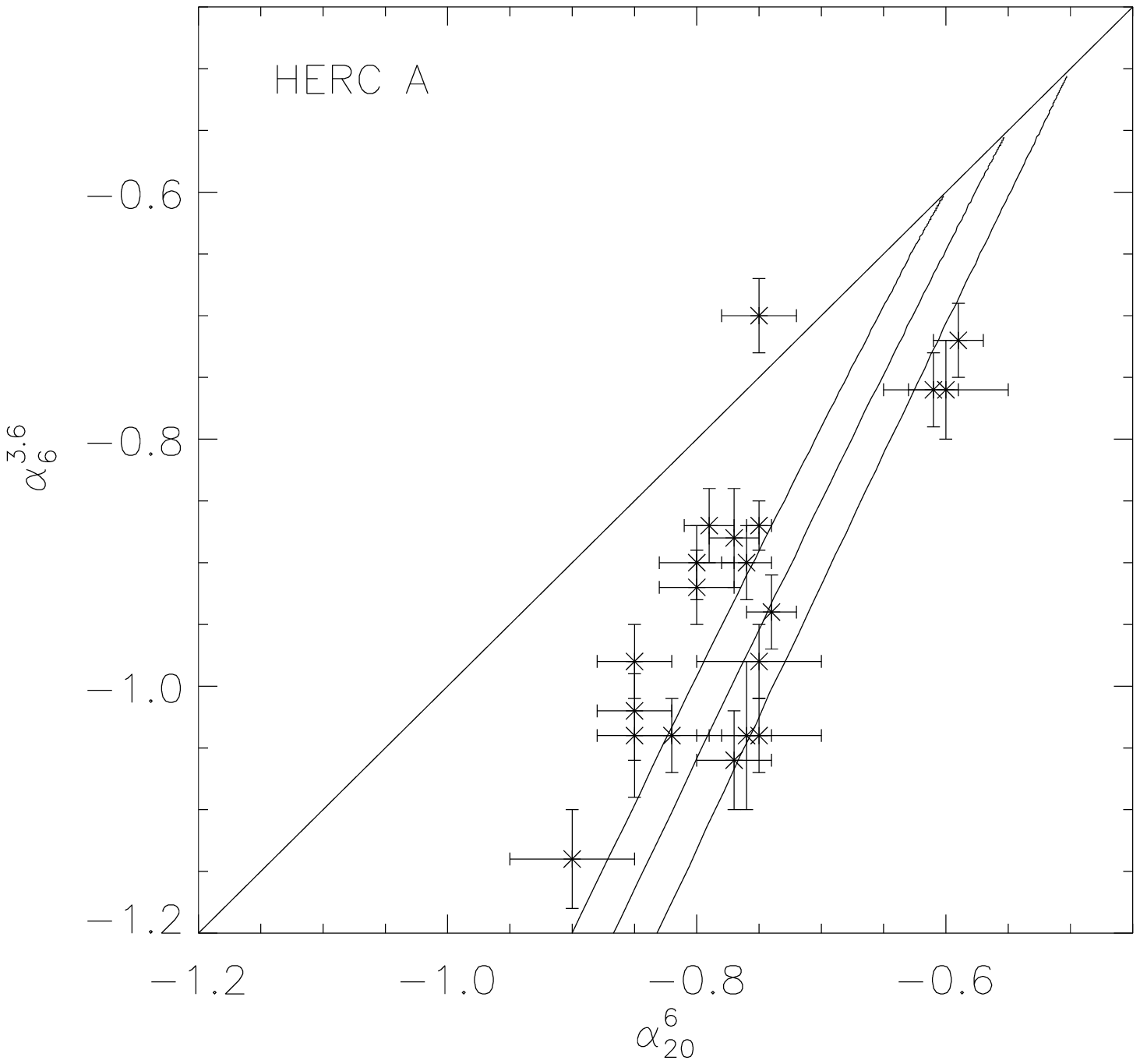}
\vspace{9pt}
\plotone{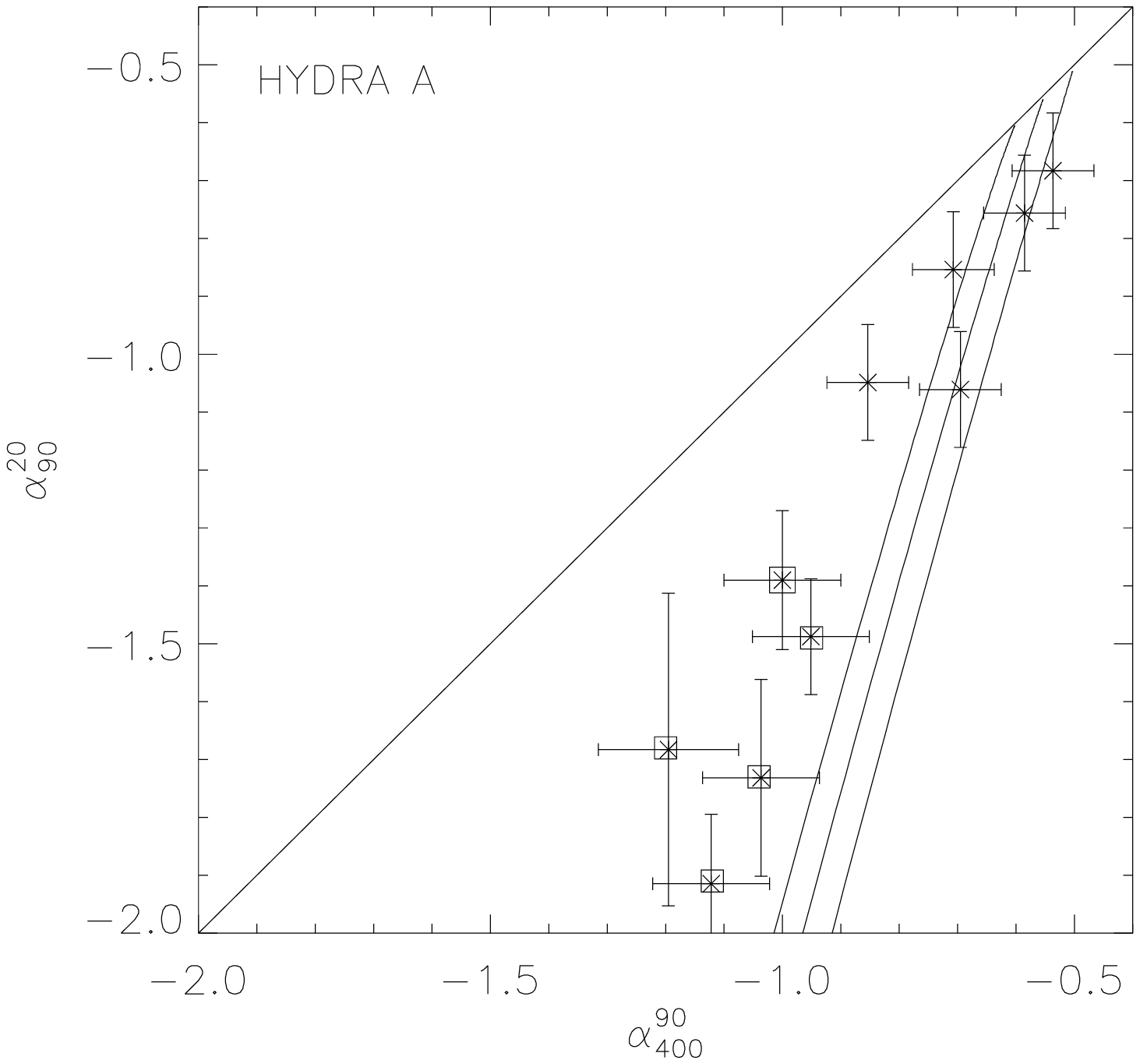}
\end{center}
\caption{Color-color diagrams:  Hercules A, using data from \citet{giz03}; 
Hydra A, using data from \citet{lane04}. Lines are as described in
Figure 11.  The fit for Hydra A is performed only for the flatter
spectrum data (crosses) before the large jump in $\alpha_{90}^{20}$. See discussion
in the text. }
\label{color2}
\end{figure}

\newpage
\begin{figure}[t]
\begin{center}
\epsscale{0.85}
\vspace{-15pt}
\hspace{-50pt}
\plotone{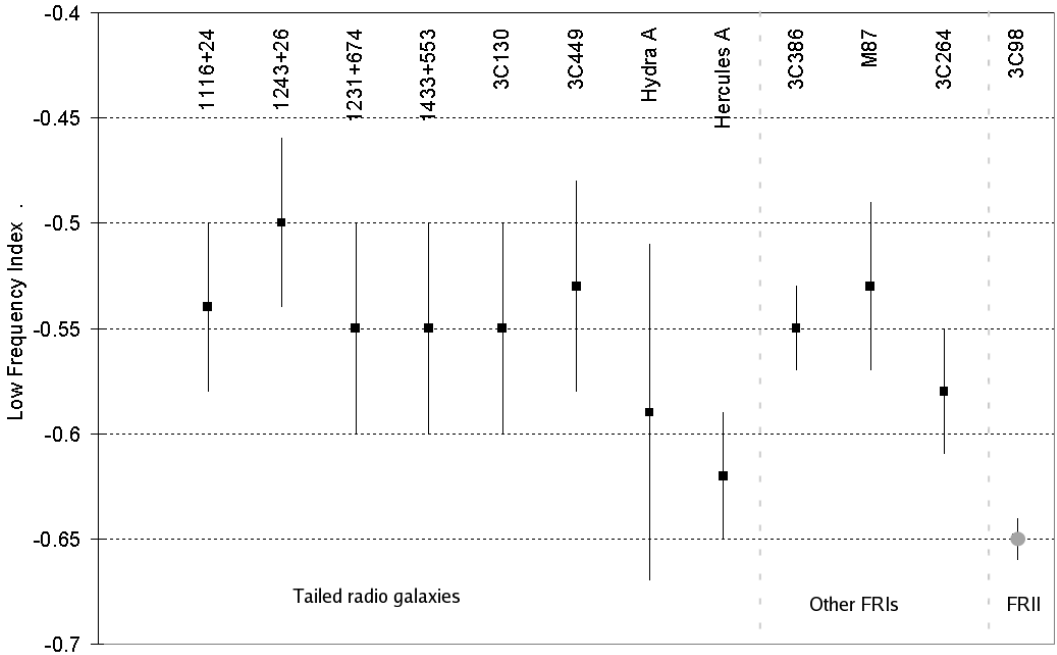}
\end{center}
\caption{Summary plot  of low frequency indices. All sources have FRI structures and luminosities except for the one FRII source, 3C98, which shows a significantly steeper low frequency index.}
\label{summary}
\end{figure}

\begin{figure}[t]
\begin{center}
\epsscale{0.4}
\plotone{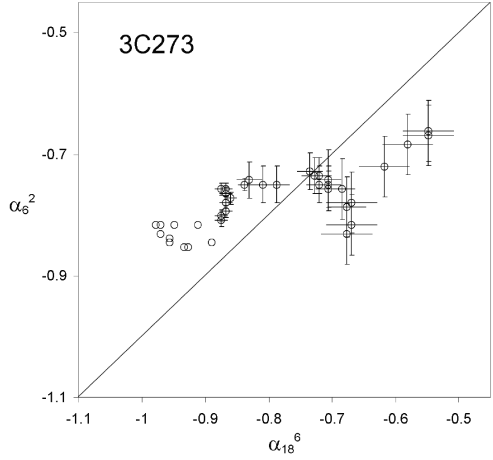}
\end{center}
\caption{Color-color diagram for 3C273, using data from \citet{con93}.
 No well-behaved single
electron population can describe this spectral shape.}
\label{273cc}
\end{figure}

\begin{figure}[t]
\begin{center}
\epsscale{0.75}
\plotone{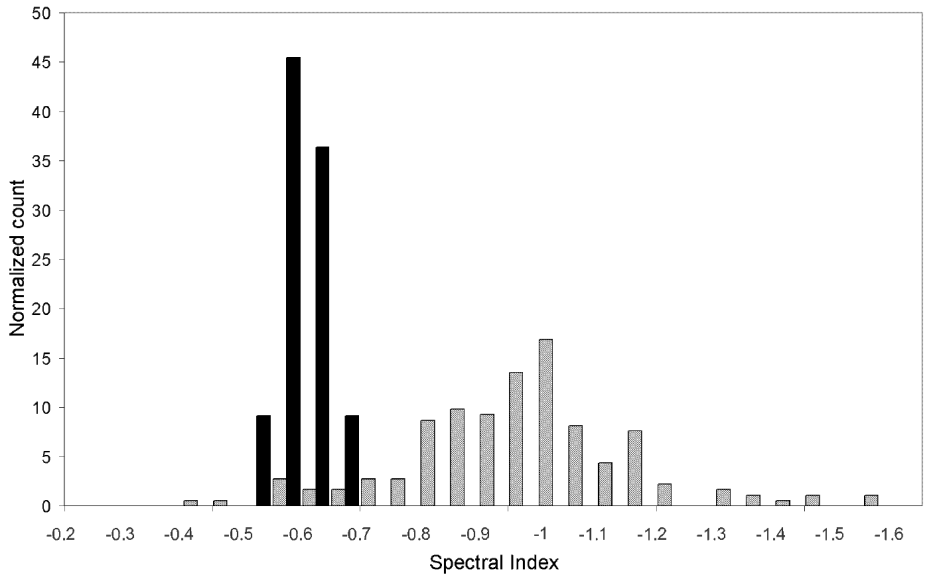}
\end{center}
\caption{Histogram of low frequency spectral indices for the sample of 11 FRI
sources described here (black columns) and the integrated spectral indices for
184  sources selected and measured by \cite{kul85a} from  the B2.3 catalog. One source with $\alpha=0.3$ is not plotted.}
\label{alphahist}
\end{figure}

\begin{figure}[t]
\begin{center}
\hspace{-60pt}
\epsscale{.65}
\plotone{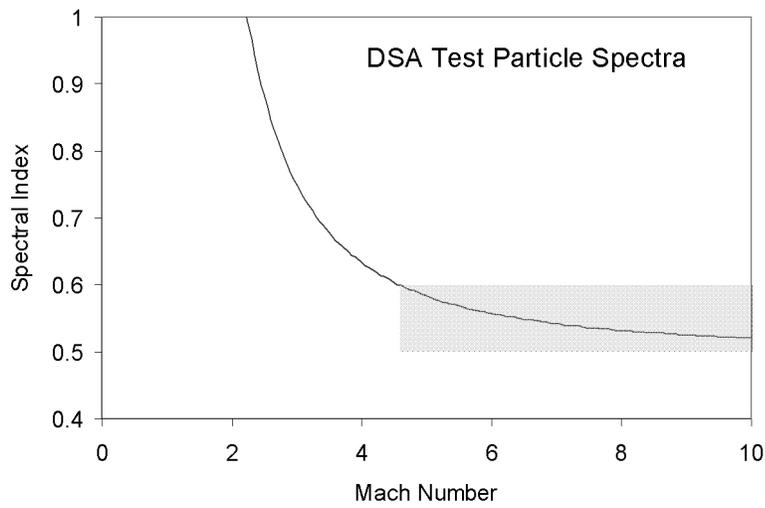}
\end{center}
\caption{Low frequency spectral index as a function of Mach number for diffusive shock acceleration in the non-relativistic test particle limit.  The grey area shows that allowed by the observations of FRIs presented here, implying a sharp lower bound to the Mach numbers for this assumed model.}
\label{alphamach}
\end{figure}

\end{document}